\documentclass[conference]{IEEEtran}
\IEEEoverridecommandlockouts
\usepackage{cite}
\usepackage{amsmath,amssymb,amsfonts}
\usepackage{algorithmic}
\usepackage{graphicx}
\usepackage{textcomp}
\usepackage{xcolor}
\usepackage{bm}

\def\BibTeX{{\rm B\kern-.05em{\sc i\kern-.025em b}\kern-.08em
    T\kern-.1667em\lower.7ex\hbox{E}\kern-.125emX}}
\begin{document}
\title{Comparison of Neural Network Architectures for Spectrum Sensing\\
\thanks{This research was supported by the Data Science/Machine Learning Initiative of UC San Diego's ECE Dept.}
\thanks{Andrew Gilman was supported by Massey University Research Fund.}
}

\author{\IEEEauthorblockN{Ziyu Ye}
\IEEEauthorblockA{\textit{Dept. ECE} \\
\textit{UC San Diego}\\
San Diego, United States \\
ziy076@kiwi-ml.ucsd.edu}\\

\IEEEauthorblockN{Kelly Levick}
\IEEEauthorblockA{\textit{Dept. ECE} \\
\textit{UC San Diego}\\
San Diego, United States \\
klevick@ucsd.edu }\and

\IEEEauthorblockN{Andrew Gilman}
\IEEEauthorblockA{\textit{Sch. of Nat. and Comp. Sci.} \\
\textit{Massey University}\\
Auckland, New Zealand \\
A.Gilman@massey.ac.nz}\\

\IEEEauthorblockN{Pamela Cosman}
\IEEEauthorblockA{\textit{Dept. ECE} \\
\textit{UC San Diego}\\
San Diego, United States \\
pcosman@eng.ucsd.edu}\and

\IEEEauthorblockN{Qihang Peng}
\IEEEauthorblockA{\textit{Sch. of Info. and Comm. Eng.} \\
\textit{UESTC}\\
Chengdu, China \\
anniepqh@uestc.edu.cn}\\

\IEEEauthorblockN{Larry Milstein}
\IEEEauthorblockA{\textit{Dept. ECE} \\
\textit{UC San Diego}\\
San Diego, United States \\
milstein@ece.ucsd.edu}
}

\maketitle

\begin{abstract}
Different neural network (NN) architectures have different advantages. 
Convolutional neural networks (CNNs) achieved enormous success in computer vision, while 
recurrent neural networks (RNNs) gained popularity in speech recognition. 
It is not known which type of NN architecture is the best fit for classification of communication signals. 
In this work, we compare the behavior of fully-connected NN (FC), CNN, RNN, and bi-directional RNN (BiRNN) 
in a spectrum sensing task. The four NN architectures are compared on their detection performance,
requirement of training data, computational complexity, and memory requirement. Given abundant 
training data and computational and memory resources, CNN, RNN, and BiRNN are shown to achieve similar 
performance. The performance of FC is worse than that of the other three types, except in the case where 
computational complexity is stringently limited. 
\end{abstract}


\section{Introduction}
A cognitive radio aims to exploit white space in existing licensed communication systems. 
Spectrum sensing, or detection of the white space, is a key focus of research in this field. 
Classic spectrum-sensing techniques, such as energy detection, matched-filter-based detection, 
cyclostationary-feature detection, and covariance-matrix-based detection were established decades ago. 
More advanced detection schemes such as cooperative sensing were proposed to further improve the 
performance by exploiting spatial, temporal and/or spectral correlations. A review of conventional 
spectrum sensing techniques can be found in \cite{yucek}. More recently, machine learning has been applied 
in spectrum sensing \cite{Bkassiny}. Our previous work \cite{peng, ye} shows potential benefits of using 
artificial neural networks (NNs) in spectrum sensing, and reviews relevant literature.

In this work, we address the choice of NN architectures for spectrum sensing. 
The two most well-studied NN architecture types today are the convolutional neural network (CNN) 
and the recurrent neural network (RNN). CNNs consist of stacked shift-invariant local filters (kernels) and are 
particularly effective in capturing spatially-local features arranged hierarchically. 
An RNN uses recurrent connections that allows it to extract and utilize empirical autocorrelations in 
sequential data. Modern RNNs are equipped with gated operations and are able to capture correlations across 
long intervals of time in the input sequence. The NN architectures have different levels of popularity in 
different tasks: Variants of CNN dominate in computer vision, while RNNs are widely used in speech recognition 
and natural language processing (NLP) . 

Various examples of NN architecture comparison for specific applications can be found in the literature. 
In \cite{zhang,yin}, CNN and RNN are compared on NLP tasks. The authors of \cite{zhang} report that 
bi-directional RNN (BiRNN) outperforms CNN in relation classification, while in \cite{yin}, CNN and RNN 
each show advantages in different tasks/scenarios. Authors of both works attribute their observations to 
the intuition that CNN puts more emphasis on local features, while RNN is able to learn long-term dependencies. 
In \cite{li}, CNN and RNN are compared on environmental-sound-based scene classification. They found that RNN 
is less effective on average, presumably because of the lack of long-term correlations in the natural 
environmental sound. In \cite{persio} and \cite{selvin},  CNN and RNN are trained to predict stock price changes. 
Both works report CNN as the winner. In \cite{liu}, RNN is found to be superior than CNN in detecting 
internet attacks based on payload data, and \cite{molina} finds that a fully-connected NN (FC) performs 
comparably to an RNN in a power grid identification task. 
 
In spectrum sensing, a preference among the NN architectures has not yet been established. 
Communication signals have both, similarities to and differences from, the types of signals considered 
in typical applications of CNN and RNN. Similar to images and speech signals, a communication signal sampled 
in time consists of ordered and correlated samples. However, it may lack some common characteristics present 
in images and speech signals. Without assuming a specific transmitted data sequence (for example, 
a known pilot signal) or a specific channel-coding scheme, it is uncertain whether the communication 
signal has strong long-term correlations, or whether it contains strong local features, such as 
corners and edges in images. Like many other applications, the choice among NN architectures 
for spectrum sensing is not obvious and requires empirical comparisons.

This work aims to provide a preliminary performance comparison of CNN, RNN, BiRNN, and FC in spectrum sensing. 
The communication signal simulation setup and the performance metrics are explained in Section II.
In Section III, we compare the best performance of each of the NN architecture types, and the 
computational complexity and memory requirements to achieve this performance.  In Section IV, 
we compare performance of NN architectures under a computational complexity constraint.
We summarize and conclude in Section V.

\section{Communication Simulation and Performance Metrics}
\subsection{Simulation Setup}
Simulation data generated with GNU Radio are used to train and test the NNs. 
Consider the following spectrum sensing task:
\begin{equation}
\bm{x} = \left\{ \begin{array}{lcl} 
\bm{s}+\bm{n} &, & y=1 \\
\bm{n} &, & y=0
\end{array} \right .
\end{equation}
where the complex vector $\bm{x}$ is the sampled low-pass equivalent of the received waveform in a 
sensing interval, and $\bm{s}$ and $\bm{n}$ are the signal and noise components of $\bm{x}$. 
$\bm{x}$ may either contain the signal component $\bm{s}$ or not, reflected by the label $y$. 
The received waveform is assumed to have been filtered by an ideal bandpass filter, whose passband matches 
the primary signal's bandwidth exactly. Parameters of the simulation are listed in Table \ref{tab_simu_params}. 
The sampling rate is set beyond the Nyquist rate of the band-limited $\bm{x}$ so that no information loss 
is incurred by sampling. Equal numbers of busy ($y=1$) and idle ($y=0$) examples are generated. 
The dataset is divided into three parts: a training set containing 8E+06 examples, and validation and test 
sets, each containing 1E+05 examples. We train different types of NNs on the training set 
to predict label $y$ from $\bm{x}$, after fine-tuning architecture and training hyperparameters on
a validation set and compare final performance on the test set. 

\begin{table}[htbp]
\caption{Communication signal simulation parameters}
\begin{center}
\begin{tabular}{|c|c|p{1.5in}|}
\hline
\textbf{Primary} & \textbf{Modulation} & \multicolumn{1}{|c|}{QPSK} \\
\cline{2-3}
\textbf{Signal} & \textbf{Pulse Shape} & root-raised-cosine (RRC) pulse with roll-off factor 0.35 \\
\cline{2-3}
& \textbf{Source Data} & random, uncorrelated bits \\
\hline
\textbf{Noise} & \multicolumn{2}{|c|}{AWGN} \\
\hline
\textbf{SNR} & \multicolumn{2}{|c|}{3dB} \\
\hline
\textbf{Sampling Rate} & \multicolumn{2}{|c|}{10 times the symbol rate} \\
\hline
\textbf{Sensing Duration} & \multicolumn{2}{|c|}{111 samples} \\
\hline
\end{tabular}
\end{center}
\label{tab_simu_params}
\end{table}

\subsection{Performance Metrics}
As suggested in \cite{canziani}, besides the detection performance, resource consumption such as 
computation and memory are also important NN performance aspects. We evaluate each NN's performance on 
four characteristics: detection performance, amount of training data required, forward-pass (inference) 
computational complexity, and forward-pass memory requirements. 

\paragraph{Detection Performance} 
The detection performance is reflected by the detection probability, $P_d$, 
which is the probability of correct decision conditional on ground truth $y=1$,
and the false alarm probability, $P_{fa}$, which is the probability of incorrect decision 
conditional on $y=0$. In this work, we fix $P_{fa}$ at $1\%$ by choosing the classifier's threshold (on the validation set), 
and evaluate the detection performance using the false dismissal probability $P_{fd}=1-P_d$.  

\paragraph{Amount of Training Data} 
In the real world, training data could be a precious resource, depending on the expense of data collection 
and labeling. The amount of training data could be limited, which may render NN models that require huge 
training sets impractical. To examine the influence of the training dataset size on performance, we train 
each NN not only on the entire 8E+06 training set, but also on subsets of size 1E+03 and 1E+05. 

\paragraph{Computational Complexity}
Computational complexity is closely related to energy consumption and decision latency, both of which
are important for spectrum sensing. The operation count, which is the number of floating point operations 
(FLOPS) in one forward pass on a single input, is used as a performance metric. We ignore the 
computation cost associated with non-linear activation functions as it is negligible in comparison 
to the number of FLOPS in the linear operations. The operation counts of FC, CNN, RNN and BiRNN 
are listed below. The symbols are defined in Table \ref{tab_denote}. 

\begin{equation}
N_{op}^{(FC)}=\sum_{k=1}^K \left(2N_{k-1}N_k\right) + 2N_K N_{out}
\end{equation} 

\begin{equation}
\begin{split}
N_{op}^{(CNN)} = &\sum_{k\in\mathcal{K}_{conv}} 2C_{k-1} C_k N_k^{(kernel)} N_k \\
&+\sum_{k\in\mathcal{K}_{pool}} C_kN_k(N_k^{(pool)}-1) \\
&+\sum_{k\in\mathcal{K}_{bn}} 2C_k N_k \\
&+ 2C_K N_K N_{dense}
+ 2N_{dense} N_{out}
\end{split}
\end{equation}

\begin{equation}
\begin{split}
N_{op}^{(RNN)} =& L \cdot \sum_{k=1}^K \left( 8(N_{k-1}+N_k) N_k + 4 N_k \right)\\
& + 2N_K N_{out}
\end{split}
\end{equation}

\begin{equation}
\begin{split}
N_{op}^{(BiRNN)} =& 2L \cdot \sum_{k=1}^K \left( 8(N_{k-1}+N_k) N_k + 4 N_k \right)\\
& + 4N_K N_{out}
\end{split}
\end{equation}

\begin{table}[htbp]
\caption{Notation}
\begin{center}
\begin{tabular}{|c|p{2in}|}
\hline
\textbf{Symbol} & \multicolumn{1}{|c|}{\textbf{Definition}} \\
\hline
$K$ & Number of hidden layers \\
\hline
$N_k$ & Layer width / feature vector length of the $k$-th hidden layer \\
\hline
$N_0$, $N_{out}$ & Input and output size \\
\hline
$N_{dense}$ & Size of the dense layer (CNN only)\\
\hline
$N_k^{(kernel)}$ & Kernel size of the $k$-th hidden layer (CNN only)\\
\hline
$N_k^{(pool)}$ & Pooling factor in the $k$-th hidden layer (CNN only)\\
\hline
$C_k$ & Layer depth / number of filters (output channels) in the $k$-th hidden layer (CNN only)\\ 
\hline
$\mathcal{K}_{conv}$ & Set of indices of convolutional layers (CNN only)\\
\hline
$\mathcal{K}_{pool}$ & Set of indices of pooling layers (CNN only)\\
\hline
$\mathcal{K}_{bn}$ & Set of indices of batch-normalization layers (CNN only)\\
\hline
$L$ & Input sequence length (RNN only)\\
\hline
\end{tabular}
\end{center}
\label{tab_denote}
\end{table}

\paragraph{Memory Requirement}
NN memory requirements depend heavily on the specific implementation. We consider two memory requirement 
metrics corresponding to two extreme cases. The peak instantaneous memory requirement, $M_{peak}$, reflects 
the peak memory requirement of the most memory-efficient implementation, which reallocates memory after 
each operation and holds only necesary content in memory at each moment. 
The total memory requirement, $M_{total}$, is the amount of memory that would be rquired if no 
reallocation happens and memory is pre-allocated for all parameters and intermediate states in advance. 
To avoid ambiguity, we assume that operations such as addition of bias term, nonlinear activation, and batch 
normalization are performed in place. We assume maximum parallelism, so operations in the same layer 
are executed in parallel. The unit of the memory metrics is the size of a floating point variable. 
The expressions for $M_{peak}$ and $M_{total}$ for FC, CNN, RNN, and BiRNN are listed below.
\begin{equation}
M_{peak}^{(FC)}=\max_k N_{k-1}\cdot N_k + 2N_k + N_{k-1}
\end{equation} 
\begin{equation}
\begin{split}
M_{peak}^{(CNN)} =&\max \left\{ M_{max}^{(conv)},  M_{max}^{(bn)}, M^{(dense)}\right\} \\ 
M_{max}^{(conv)} =& \max_{k\in\mathcal{K}_{conv}} \{C_{k-1}N_{k-1} + C_k N_k \\ &+ C_k (C_{k-1} N_k^{(kernel)} + 1) \}\\
M_{max}^{(bn)} =& \max_{k\in\mathcal{K}_{bn}} 3C_k N_k  \\
M^{(dense)} =& C_K N_K N_{dense} + C_K N_K + 2N_{dense}
\end{split}
\end{equation} 
\begin{equation}
M_{peak}^{(RNN)}=\max_k  4(N_{k-1}+N_k)N_k+N_{k-1}+6N_k
\end{equation} 
\begin{equation}
M_{peak}^{(BiRNN)}=2M_{max}^{(RNN)}
\end{equation} 
\begin{equation}
\begin{split}
M_{total}^{(FC)} = & N_0 + \left(\sum_{k=1}^K (N_{k-1}\cdot N_k +2N_k)\right)\\ & + (N_K\cdot N_{out} + 2N_{out})
\end{split}
\end{equation}
\begin{equation}
\begin{split}
& M_{total}^{(CNN)} = C_0 N_0\\
& 
+\sum_{k\in\mathcal{K}_{conv}} \left( C_k (C_{k-1}N_k^{(kernel)}+1) + C_kN_k \right) \\
&+\sum_{k\in\mathcal{K}_{pool}} \left( C_{k-1}N_{k-1}+C_kN_k \right) +\sum_{k\in\mathcal{K}_{bn}} 2C_kN_k\\
&+\left( C_K N_KN_{dense}+2N_{dense} \right)
+\left(N_{dense}N_{out} +2N_{out}\right)
\end{split}
\end{equation}
\begin{equation}
\begin{split}
M_{total}^{(RNN)} = & N_0 + \sum_k \left( 4(N_{k-1}+N_k)N_k  +10N_k \right)\\ 
& + (N_KN_{out} + 2N_{out})
\end{split}
\end{equation}
\begin{equation}
\begin{split}
M_{total}^{(BiRNN)} = & N_0 + 2\sum_k \left( 4(N_{k-1}+N_k)N_k+10N_k \right)\\ 
&  + (2N_KN_{out} + 2N_{out})
\end{split}
\end{equation}

\section{Comparison of Optimized NNs} \label{sec_comparison}
As performance is multi-dimensional, ideally, the performance of each NN architecture type should be 
characterized by the Pareto frontier in the multi-dimensional space. To obtain one point on the Pareto frontier 
requires solving a constrained optimization problem, in which the training data size, 
computational complexity and memory requirement are fixed, and the architecture and learning hyperparameters are optimized to maximize the 
detection probability. Due to limited time and computational resources, instead of searching for the Pareto 
frontier, we resolved to the following protocol: For each of the four NN architecture types, the hyperparameters are manually tuned 
to maximize the detection probability on three training set sizes (1E+03, 1E+05, and 8E+06). 
Complexity and memory were ignored during this tuning process, 
but are compared on the best set of hyperparameters. 

In tuning the hyperparameters, the following constraints are imposed. For CNN, we consider 
a type of architecture adopted from VGGNet \cite{simonyan}. The CNN consists of multiple
homogeneous blocks in a sequence, followed by a single dense layer. Each block consists of two
convolutional layers (kernel size=3, stride=1, padding=`same' mode, ReLU activations) alternating with
batch normalization layers and followed by a max-pooling layer with a small pooling factor (2-4).
Both convolutional layers within a block contain the same number of channels and the number of channels
increases by a factor of 2 in each consecutive block.
The output of the last block is flattened into a vector and passed to a dense 
layer of the same size.  The output of the dense layer is fed to the output layer.
For RNN, we confine our search within the long-short-term-memory (LSTM) architecture. 
Because VGG and LSTM are state-of-the-art models in the CNN and RNN families respectively, 
we believe that these constraints do not cause much performance loss in detection probability.

We manually optimized the following hyperparameters to the best of our ability: the number of hidden layers (FC and RNNs only),
the size of hidden layers (FC and RNNs only), number of blocks (CNN only), the pooling factor (CNN only), the learning rate, and the batch size (see Table \ref{tab_params} for details). 
Adam optimizer with learning-rate scheduling and early termination are used for all training.
The learning rate is reduced by a factor of 10 each time when the validation loss sees no notable 
decrease in 10 consecutive epochs.  Early termination is triggered if the validation loss sees no 
notable decrease in 15 consecutive epochs. Due to the stochastic nature of initialization and training, 
we repeat training on the two smaller training sets 10 times, and 5 times on the largest training set. 
The maximum and median detection probabilities are then computed.

\begin{table}[htbp]
\caption{Tuned hyper-parameters}
\begin{center}
\begin{tabular}{|p{0.25in}|p{0.4in}|p{0.4in}|p{0.3in}|p{1.2in}|}
\hline
\textbf{Arch. Type}  	&\textbf{Training Data Size}		&\textbf{Learning Rate}	&\textbf{Batch Size} &\textbf{Model Specification} 	\\
\hline
FC		& 1E+03			&1e-3&20		& 4 hidden layers, each of size 64\\
\cline{2-5}
		& 1E+05			&5e-4&1000	& Same as above\\
\cline{2-5}
		& 8E+06			&5e-4&1000	& Same as above\\
\hline
CNN	& 1E+03			&1e-3&1000	& 2 blocks with 32, 64 filters respectively. Pooling factor is 4. \\
\cline{2-5}
		& 1E+05			&5e-4&1000	& Same as above\\ 
\cline{2-5}
		& 8E+06			&5e-4&1000	& 3 blocks with 16, 32, 64 filters respectively. Pooling factor is 2.\\
\hline
RNN	& 1E+03			&1e-4&50		& 1 hidden layer of size 64\\
\cline{2-5}
		& 1E+05			&5e-4&100		& 1 hidden layer of size 128\\
\cline{2-5}
		& 8E+06			&5e-4&100  	& Same as above\\
\hline
BiRNN	& 1E+03			&5e-4&50		& 1 hidden layer of size 64\\
\cline{2-5}
		& 1E+05			&5e-4&1000	& Same as above\\
\cline{2-5}
		& 8E+06			&5e-4&1000	& 1 hidden layer of size 128\\
\hline
\end{tabular}
\end{center}
\label{tab_params}
\end{table}

\begin{figure}[htbp]
\centerline{\includegraphics[width=3.7in]{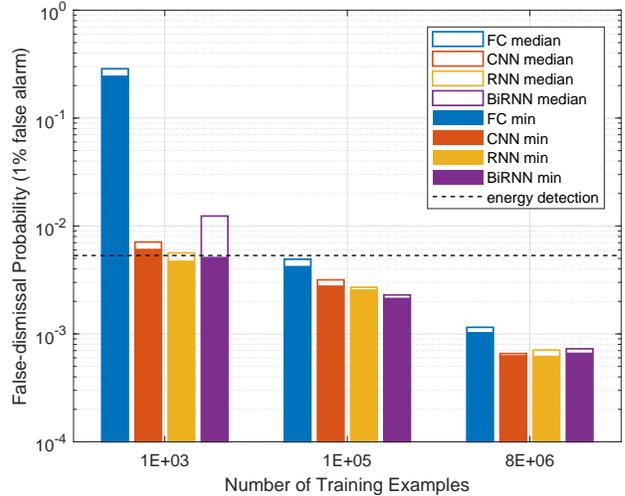}}
\caption{False dismissal probabilities of the optimized NNs tested on SNR=3dB, QPSK test data}
\label{fig_p_d}
\end{figure}

The false dismissal probabilities of the NNs with tuned hyperparameters are compared in Fig.~\ref{fig_p_d}. 
The FC gives the worst performances among the four architecture types across all of the training set sizes. 
The FCs trained on the two smaller training sets perform either worse than or similar to the energy detection, 
and have no practical value considering that the energy detection is much simpler and requires no training. 
Unlike CNN, RNN or BiRNN, the FC does not observe the inherent order and adjacency of samples in the input, 
which makes it much harder to learn local patterns, no matter how wide/deep the FC is. 
The performances of the CNN, RNN and BiRNN are notably better than that of the FC, and can be considerably better than 
energy detection when the training set is large enough. The false dismissal probabilities of the CNN, RNN 
and BiRNN trained on the size-8E+06 training set are lower than that of the energy detection by at least 
a factor of 8. The performance gaps between CNN, RNN, and BiRNN are less defined, 
and not consistent across the training set sizes. Considering that our hyperparameter tuning is relatively 
coarse, no distinction can be comfirmed between CNN, RNN and BiRNN's abilities in this detection task.
 
\begin{figure}[htbp]
\centerline{\includegraphics[width=3.7in]{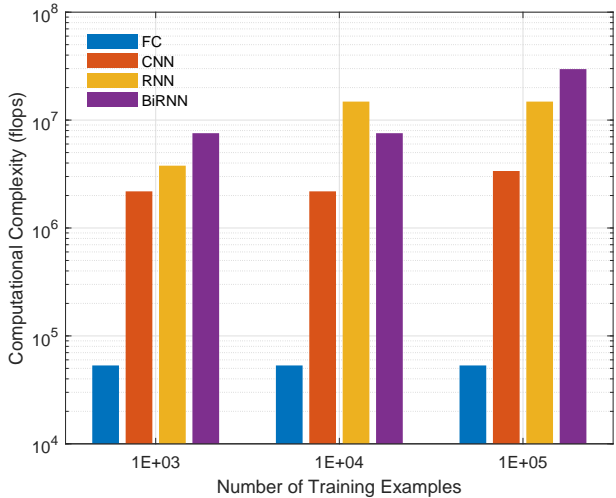}}
\caption{Operation counts of the optimized NNs}
\label{fig_comp_cplx}
\end{figure}

\begin{figure}[htbp]
\centerline{\includegraphics[width=3.7in]{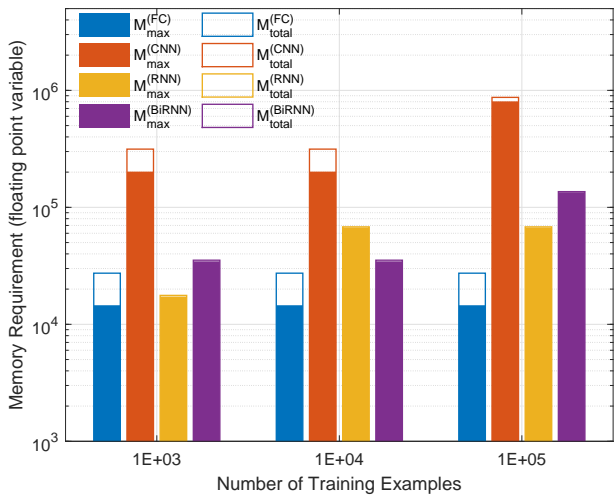}}
\caption{Memory requirements of the optimized NNs}
\label{fig_mem}
\end{figure}

The operation counts and memory requirements of the tuned NNs are shown in Figs.~\ref{fig_comp_cplx} 
and \ref{fig_mem}.  These metrics were not taken into consideration in the hyperparameter tuning process, so
the tuned NNs could be highly inefficient in computation and memory. It is likely that by carefully 
changing the hyperparameters, the operation counts and memory requirements shown in Figs.~\ref{fig_comp_cplx} 
and \ref{fig_mem} could be somewhat reduced without impairing detection performance. 
Despite these limitations, some qualitative obervations can be made from the figures. 
First, in Fig.~\ref{fig_comp_cplx}, the operation count of the FC is lower than that of the other NNs by 2 
orders of magnitude on average. While the tuned FC gives the worst detection performance in Fig.~\ref{fig_p_d}, 
it also consumes the least amount of computational resources. Inspired by this observation, 
further comparisons between the FC and the other NN architecture types are conducted in 
Section \ref{sec_comparison_under_constraint}. Second, while the RNNs and BiRNNs require more 
computation than the CNNs (Fig.\ref{fig_comp_cplx}), their memory requirements are notably lower than 
that of the CNNs (Fig.~\ref{fig_mem}). Considering that the memory for storing the parameters constitutes 
most of the total memory, this observation likely implies that the RNN has a higher level of 
parameter sharing than the CNN, which would potentially make the RNN more memory efficient. 
On the 1E+05 training set, the RNN requires more computation and memory than the BiRNN.  This is because 
the RNN tuned on the 1E+05 training set has a larger layer size than the BiRNN tuned on the same training set.

\section{Comparison under a Computation Constraint} \label{sec_comparison_under_constraint}
\begin{figure}[htbp]
\centerline{\includegraphics[width=3.7in]{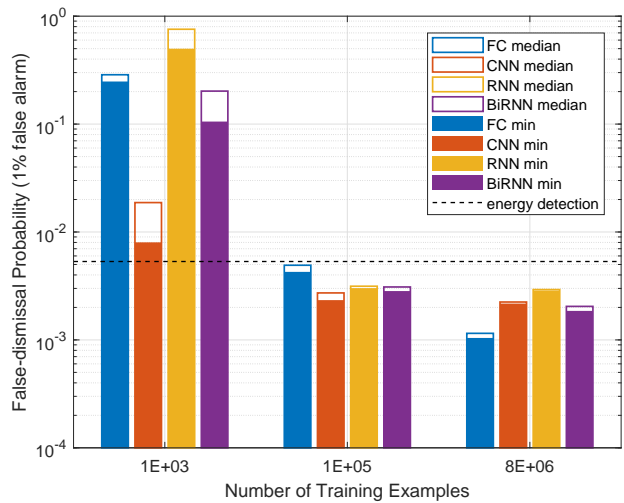}}
\caption{False dismissal probability of NNs under computational complexity constraint}
\label{fig_p_d_2}
\end{figure}

In the previous section, we compared the NNs whose hyperparameters were tuned with no constraint on 
their computational complexity or memory. The comparison shows a large gap between the computational 
complexity of the FC and that of the other NNs. While the FC clearly has the disadvantage that it was 
not able to approach the same level of performance of the other NNs in the hyperparameter tuning process, 
it is unclear how its performance compares to CNN, RNN and BiRNN with the same computational complexity. 
We scaled down the CNN, RNN and BiRNN used in the previous section so that their operation counts roughly match 
that of the FC. The modified NNs are described in Table \ref{tab_params_modified}. 
The four NNs with roughly the same level of computational complexity are trained on the three training sets, 
and their false dismissal probabilities are plotted in Fig.~\ref{fig_p_d_2}.  The CNN, RNN, and BiRNN perform
worse through the modification, particularly on the largest training set, where they all perform worse than the FC. 

Because of the limitations of our parameter tuning process, the NNs which gave the performances in 
Fig.~\ref{fig_p_d_2} are not necessarily the most computationally efficient in the CNN, RNN and BiRNN families. 
In particular, the constraints we imposed on the CNN and RNN architectures in Section \ref{sec_comparison} could 
have limited their computational efficiency. It is highly possible that there exist some CNN, RNN and BiRNN which 
can achieve lower false dismissal probability without violating the computation constraint. 
However, the observation in Fig.~\ref{fig_p_d_2} suggests that unlike in the general case where the FC 
is clearly worse than the other architecture types, there is a possibility that an FC can achieve 
a performance comparable to that of the more advanced architectures when the computational resources are 
stringently limited.

\begin{table}[htbp]
\caption{hyper-parameters modified to meet the computational complexity constraint}
\begin{center}
\begin{tabular}{|p{0.5in}|p{2.5in}|}
\hline
\textbf{Arch. Type}  	&\textbf{Model Specification} 	\\
\hline
FC								& 4 hidden layers, each of size 64 \\
\hline
CNN							& 1 CONV-CONV-POOL block with 4 filters. Pooling factor is 4. \\
\hline
RNN							& 1 hidden layer of size 6\\
\hline
BiRNN							& 1 hidden layer of size 4\\
\hline
\end{tabular}
\end{center}
\label{tab_params_modified}
\end{table}

\section{Conclusion}
In this work, we compared detection performance, computational complexity, and memory requirements 
of four different types of NN architectures in a spectrum sensing task. We found that with abundant 
computation and memory resources, CNN, RNN and BiRNN are able to achieve a performance significantly 
better than that of the energy detector. CNN, RNN and BiRNN architectures resulted in very similar detection performance. 
The RNN/BiRNN possibly have an advantage over the CNN in terms of memory efficiency. 
Experimental results also show that FC should not be used in spectrum sensing unless in the special 
case where the computational resource is stringently limited. 
One factor not considered in this work is the correlation in the data carried by the signal, 
which commonly exists because of the error correction coding and correlations in the source content. 
The effect of these correlations on detection performance of the NN architectures 
is a potential topic of future research.

\end{document}